\documentclass[a4paper]{article}

\usepackage{tikz}
\usetikzlibrary{positioning}
\usetikzlibrary{backgrounds}
\usetikzlibrary{fit}
\usetikzlibrary{arrows.meta}
\usepackage{pgfplots}
\pgfplotsset{compat=1.13}
\usepackage{longtable}
\usepackage{rotating}
\usepackage{standalone}

\usepackage[most]{tcolorbox}

\newtcolorbox{greybox}{
  colback=gray!10,
  colframe=gray!50,
  boxrule=0.4pt,
  arc=1mm,
  left=7pt,
  right=7pt,
  top=7pt,
  bottom=7pt
}

\usepackage{amssymb}
\usepackage{array}
\usepackage{booktabs}
\usepackage{siunitx}
\usepackage{graphicx}
\usepackage{graphbox}
\usepackage{hyperref}
\hypersetup{
colorlinks=true,
citecolor=black,
linkcolor=black,
urlcolor=black
}
\usepackage{tabularx}

\newcommand{\attfact}{attenuation factor}
\newcommand{\af}{\ensuremath{F}}

\newcommand{\msbe}[2]{\ensuremath{\mathit{MSBE}^{(\mathit{#1},\, \mathit{#2})}}}

\usepackage{xcolor}

\usepackage{vwcol}

\title{Auto-adaptive Resonance Equalization using Dilated Residual Networks}
\author{Maarten Grachten$^1$ and Emmanuel Deruty$^2$ and Alexandre Tanguy$^3$}

\date{}

\begin{document}

\maketitle{}

\begin{center}
  $^1$Contractor for Sony CSL Paris, France\\ $^2$Sony CSL Paris, France\\ $^3$Yascore, Paris, France
\end{center}

\vspace{1cm}

\begin{greybox}
\begin{small}
Published version: Grachten, M., Deruty, E. and Tanguy, A. (2019).
Auto-adaptive Resonance Equalization using Dilated Residual Networks.
\textit{Proceedings of the 20th International Society for Music Information Retrieval Conference}, 405--411.
Delft, The Netherlands. \\\url{http://archives.ismir.net/ismir2019/paper/000048.pdf}
\end{small}
\end{greybox}

\vspace{1cm}

\begin{abstract}
In music and audio production, attenuation of spectral resonances is an important step towards a technically correct result.
In this paper we present a two-component system to automate the task of resonance equalization.
The first component is a dynamic equalizer that automatically detects resonances and offers to attenuate them by a user-specified factor.
The second component is a deep neural network that predicts the optimal attenuation factor based on the windowed audio.
The network is trained and validated on empirical data gathered from an experiment in which sound engineers choose their preferred attenuation factors for a set of tracks.
We test two distinct network architectures for the predictive model and find that a dilated residual network operating directly on the audio signal is on a par with a network architecture that requires a prior audio feature extraction stage.
Both architectures predict human-preferred resonance attenuation factors significantly better than a baseline approach.
\end{abstract}

\newpage

\section{Introduction and related work}\label{sec:introduction}

Equalization is part of the audio mixing and mastering process.
It is a redistribution of the energy of the signal in different frequency bands.
The process has been traditionally performed by skilled sound engineers or musicians who determine the proper equalization given the character and peculiarities of the input audio.
In recent years, methods have been developed for semi-automatic and automatic equalization.
Such methods may be used by audio professionals to save time, or by recording enthusiasts lacking the skills required to use manual equalization tools effectively.
These methods include automatic detection of frequency resonances~\cite{bitzer2009automatic},
automatic equalization derived from expert practices~\cite{de2013knowledge}, and
automatic conformation to a target spectrum~\cite{ma2013implementation}.
Equalization profiles may also be derived from semantic descriptors~\cite{cartwright2013social}.
Appropriate equalization settings can be found through different means, for example by comparing the input source to previously equalized content~\cite{Reed:2000:PAS:325737.325848}, or by formulating equalization as an optimization problem in which inter-track masking is used as the cost function~\cite{hafezi2015autonomous}.
Some automated equalization functionalities are featured in commercial products.
Examples include the "learn" function in Neutron 2's equalizer\footnote{\scriptsize \url{https://www.izotope.com/en/products/mix/neutron/neutron-advanced.html}}, and the "Tame" function in SoundTheory's Gulfoss\footnote{\scriptsize \url{https://www.soundtheory.com/home}}.

The use of machine learning to solve audio production related tasks is recent.
Aside from an early approach using nearest neighbor inference to infer equalization~\cite{Reed:2000:PAS:325737.325848}, most applications of machine learning for automatic mixing date from the past few years.
Automatic mixing tasks that have been addressed in this way include automatic reverbation~\cite{chourdakis2017a}, dynamic range compression~\cite{mimilakis2016deep}, and demixing/remixing of tracks~\cite{mimilakis16:_new_sonor_jazz_recor}.
All three studies use neural networks, which are rapidly becoming a \textit{de facto} standard method for machine learning.
To our knowledge, there is no documented example of the use of neural networks for automatic equalization.

A particular method of equalization is the attenuation of resonating or salient frequencies, \textit{i.e.} frequencies that are substantially louder than their neighbors~\cite{bitzer2008evaluating}.
The focus of this paper is the automation of such as process using a deep neural network.
Salient frequencies may originate from different phenomena, such as the acoustic resonances of a physical instrument or an acoustic space.
They may be considered a deficiency, in the sense that they may mask the content of other frequency regions.
One particular difficulty in resonance attenuation is finding the right amount of attenuation.
For example, too much attenuation may unmask noise that would otherwise remain unheard, or flatten the spectrum to the point of garbling the original audio.

Our method fully automates the resonance attenuation process.
It includes, 1) a windowed, dynamic resonance attenuation process that works on 0.5s windows and can be controlled with a single parameter---the \emph{attenuation factor}, 2) a deep neural network that predicts the attenuation factor from the input audio, making the process auto-adaptive~\cite{reiss2018applications}.

\newpage

For the training and validation of the predictive model we conduct a listening experiment determining optimal resonance attenuation factors for a set of tracks, as chosen by sound engineers.
We describe and test two alternative modeling approaches, and find that a state-of-the-art convolutional network for image processing can be successfully adapted for audio processing.
Experimental results show that this network architecture, which directly processes the raw audio signal, is on a par with a more traditional approach of training a neural network on a set of pre-computed audio descriptors.

The paper is organized as follows.
Section~\ref{sec:reson-equal} describes the resonance equalization process.
The listening experiment is described in Section~\ref{sec:listening-experiment}.
The design, training, and evaluation of the predictive models is presented in Section~\ref{sec:learn-optim-atten}, and conclusions are presented in Section~\ref{sec:conclusion}.

\section{Resonance Equalization}\label{sec:reson-equal}

Traditionally, resonance attenuation has been a manual task in which a musician or sound engineer determines the resonating frequencies by ear or using a graphical tool, in order to reduce the energy of the signal in those frequencies by an appropriate amount~\cite{mccandless2017craft}.
In this section, we describe a procedure that identifies resonating frequencies autonomously, and reduces the energy in those frequencies by a factor that is controlled by the user.
The procedure works on overlapping audio windows that must be large enough to allow for spectral analysis at a high frequency resolution.

\begin{figure}
\centering
\begin{tikzpicture}[
  node distance = .6cm
  ]
  \newcommand{\minp}{6em}
  \newcommand{\mins}{12em}
  \newcommand{\minst}{10em}
  \newcommand{\smins}{9em}
  \tikzstyle{proc} = [minimum width=\mins, fill=gray!20, text width=\minst, align=center]
  \tikzstyle{data} = [draw, minimum width=\mins, fill=gray!0, text width=\minst, align=center]
  \tikzstyle{param} = [draw, minimum width=7em, fill=green!30, text width=5em, align=center]
  \tikzstyle{const} = [minimum width=6em, fill=blue!30, text width=5em, align=center]
  \tikzstyle{narrow} = [minimum width=8em, text width=7em]
  \newcommand{\mynode}[4]{\node[#1] (#2) {\rule{0em}{2ex}#3}; \node[gray,anchor=north west, at=(#2.north west)]{#4};}
  \mynode{data}{input}{PCM Signal}{a};
  \mynode{proc, below= of input}{fft}{DFT}{b};
  \mynode{data, below= of fft}{spec}{Power Spectrum}{c};
  \mynode{proc, below= of spec}{elc}{Equal Loudness Weighting}{d};
  \mynode{data, below= of elc}{ps}{ELC-weighted Power Spectrum}{e};
  \mynode{proc, below= of ps}{smooth}{Smoothing}{f};
  \mynode{data, below= of smooth}{sps}{Smoothed Power Spectrum}{g};
  \mynode{proc, narrow, right= of sps}{pdiff}{Positive Difference}{h};
  \mynode{proc, narrow, right= of spec}{mult}{$\odot$}{p};
  \mynode{data, right= of mult}{cspec}{Corrected Power Spectrum}{q};
  \mynode{proc, above= of cspec}{ifft}{inverse DFT}{r};
  \mynode{data, above= of ifft}{cinput}{Corrected PCM Signal}{s};
  \mynode{data, narrow, right= of elc}{corr}{Spectrum Correction}{o};
  \mynode{proc, right= of corr}{ielc}{Log-to-linear/freq-interpolation}{n};
  \mynode{data, below= of ielc}{pscorr}{Power Spectrum Correction}{m};
  \mynode{proc, below=.4 of pscorr}{smult}{$*$}{k};
  \mynode{proc, below=.2 of smult}{neg}{$-$}{j};
  \mynode{param,above left=0.2cm and .6cm of smult}{F}{Attenuation Factor}{l};
  \mynode{data, right= of pdiff}{res}{Resonance}{i};


  \foreach \from/\to in {
    input/fft,
    fft/spec,
    spec/elc,
    elc/ps,
    ps/smooth,
    smooth/sps,
    ps.east/pdiff,
    sps/pdiff,
    spec/mult,
    cspec/ifft,
    mult/cspec,
    corr/mult,
    F/smult.west,
    pdiff/res,
    res/neg,
    neg/smult,
    smult/pscorr,
    pscorr/ielc,
    ielc/corr,
  ifft/cinput}
  \draw [->] (\from) -- (\to);
  
\end{tikzpicture}
\caption{Resonance equalization block diagram; White and gray blocks represent data and processes respectively; The green block depicts the single user-controlled parameter; The symbols $\odot$, $*$ and $-$ represent elementwise vector/vector multiplication, elementwise scalar/vector multiplication, and unary negation respectively.}
\label{fig:reseq}
\end{figure}
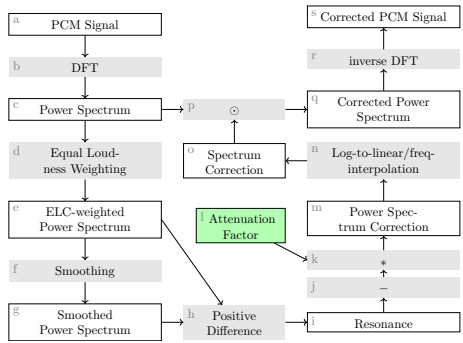

Figure~\ref{fig:reseq} displays a block diagram of the resonance attenuation process, where each element is denoted by a letter.
In the following, we will use these letters to refer to the corresponding elements in the diagram.
First the audio signal is used to compute a power spectrum weighted by Equal-Loudness Contours (ELC)~\cite{iso2003226} at a fixed monitoring level of 80 phon (Figure~\ref{fig:reseq}, element d) in order to reflect the perceptual salience of the signal energy at different frequency bands.
The value of 80 phon is chosen in relation to the procedure detailed in Section \ref{sec:firstprocedure}.
The ELC-weighted power spectrum (e) consists of 400 log-scaled frequency bands.

Resonances (i) are determined by smoothing ELC-weighted power spectrum (e) to obtain (g) and computing the elementwise differences (e) minus (g), setting negative elements to zero (h).
The negative of the resonances is then scaled by the user defined attenuation factor (l), transformed back to a linear scale and converted back to the shape of the original spectrum using interpolation (h).
The result (o) is a vector of scaling factors (one for each DFT bin) that range between 0 and 1 for resonating frequencies and are 1 otherwise.
Multiplying the original power spectrum (c) with the scaling factors gives the corrected power spectrum (q) from which the corrected audio signal (s) is recovered through the inverse DFT (r).

\newpage

\section{Listening experiment}\label{sec:listening-experiment}
A listening experiment was carried out to obtain ground truth in terms of optimal resonance attenuation values for a set of audio tracks.
In the experimental design of the listening test it proved unpractical to ask subjects to set a varying resonance attenuation factor.
Therefore, we chose relatively homogeneous sound fragments, and let the subjects choose a single attenuation factor for the whole fragment.

\subsection{Participants}
A group of 15 subjects was recruited for the experiment, in the area of Paris (France).
All of them are recognized professionals in the industry.
Nine subjects specialize in studio recording (Classic, Jazz, Pop, Rock, Movie Music, audio post-production), three are experts in live music, and three are composer/music producer.
The subjects were between 24 and 42 years old, with an average age of 32 years.
They were recruited and paid as if they were working on a commercial project.

\newpage

\subsection{Data}
A set of 150 audio tracks was used for the listening experiment.
The tracks are excerpts from longer pieces, with a mean duration of 46 seconds and a standard deviation of 16 seconds.
All tracks were processed using Nugen AMB R128\footnote{\scriptsize \url{https://nugenaudio.com/amb}} so that they were aligned to the same median loudness.
The set comprised pop and rock music, as well as film scores.
Of this set, 131 tracks were unique recordings, while the remaining 19 tracks were variants of some of the unique 131 recordings, with differences in mixing.
None of the tracks were previously mastered.

\subsection{Procedure}\label{sec:firstprocedure}
The listening experiment took place in a recording studio, where participants listened to the audio tracks individually, using studio monitors, at a measured listening loudness of 80 dBC.
This value was chosen as being representative of a normal listening loudness during audio production.
The participants were presented with a web interface in which they could listen to each track with different degrees of resonance attenuation, ranging from 0 (no attenuation) to 1 in 17 steps.
They could select their preferred degree of resonance attenuation, or alternatively decline to select any version, indicating that none of the versions sounded acceptable.
The tracks were separated from each other by 10 seconds of pink noise surrounded by a short silence to give the participants a fixed reference.
Sessions of 50 tracks were alternated with breaks.

\subsection{Results and discussion}
\begin{figure}
\centering
\includegraphics[width=\linewidth]{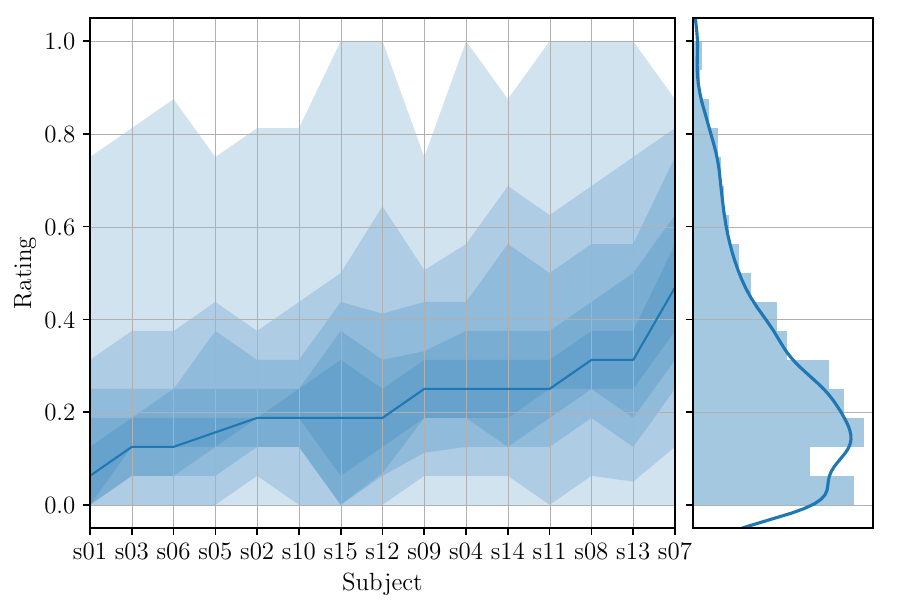}
\caption{Rating percentiles (in steps of 10\%) per subject. Darker areas correspond to more central percentile ranges, lighter areas to more peripheral ranges. The bold line in the left plot shows the median ratings.}
\label{fig:pp_subject}
\end{figure}

Basic statistics of the results per subject are given in Table~\ref{tab:ratingstats}.
Subject 13 stands out because of the number of missing ratings (21 versus a median of 1 over all subjects).
Subjects 1 and 15 have abnormally high rates of 0.0 ratings (72 and 58 respectively, versus a median of 16 over all subjects).
Finally, Subject 7 stands out in terms of median rating (0.469 versus a median of 0.188 over all subjects).
Figure~\ref{fig:pp_subject} shows the distribution of ratings per subject.

\begin{table}
\caption{Rating statistics per subject.
Outlying values are highlighted in bold (see text).}
\label{tab:ratingstats}
\small
\centering
\begin{tabular}{crrrrr}
\toprule
\multicolumn{1}{c}{Subject} & \multicolumn{1}{l}{\# No rating} & \multicolumn{1}{c}{\# 0.0} & \multicolumn{1}{c}{Min} & Median & Max \\ \midrule
s01 &  1 &  \textbf{72} &  0.0 &  0.062 &  0.750 \\
s02 &  0 &   2 &  0.0 &  0.188 &  0.812 \\
s03 &  2 &  25 &  0.0 &  0.125 &  0.812 \\
s04 &  2 &   7 &  0.0 &  0.250 &  1.000 \\
s05 &  4 &  25 &  0.0 &  0.156 &  0.750 \\
s06 &  0 &  22 &  0.0 &  0.125 &  0.875 \\
s07 &  0 &   2 &  0.0 &  \textbf{0.469} &  0.875 \\
s08 &  8 &   9 &  0.0 &  0.312 &  1.000 \\
s09 &  0 &   9 &  0.0 &  0.250 &  0.750 \\
s10 &  1 &  17 &  0.0 &  0.188 &  0.812 \\
s11 &  1 &  16 &  0.0 &  0.250 &  1.000 \\
s12 &  2 &  25 &  0.0 &  0.188 &  1.000 \\
s13 & \textbf{21} &  13 &  0.0 &  0.312 &  1.000 \\
s14 &  0 &   3 &  0.0 &  0.250 &  0.875 \\
s15 &  0 & \textbf{58} &  0.0 &  0.188 &  1.000 \\
\bottomrule
\end{tabular}
\end{table}

\begin{figure}
\centering
\includegraphics[width=1.0\linewidth]{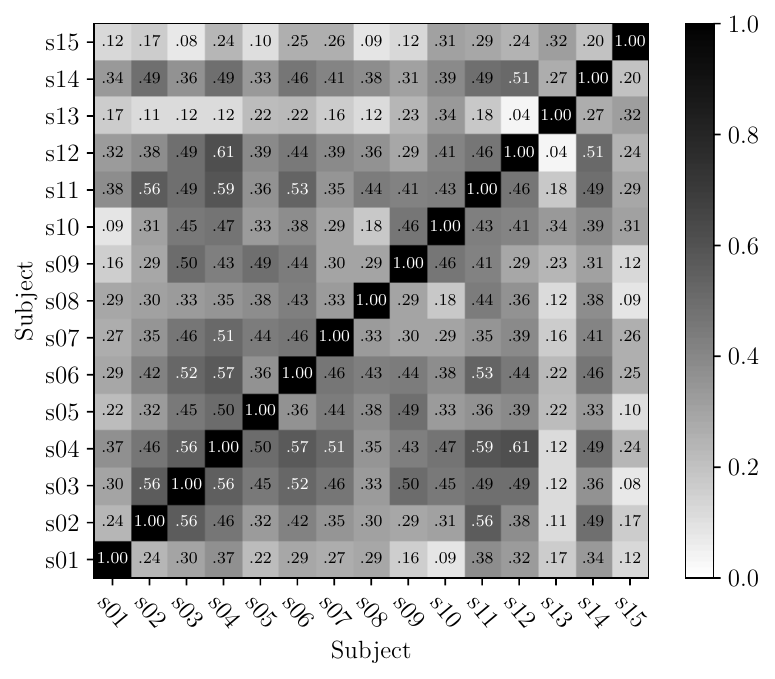}
\caption{Correlation coefficients of ratings among subjects.}
\label{fig:cc_subjects}
\end{figure}

To see how strongly the ratings are linearly related among subjects, we compute the Pearson correlation for each pair of subjects (Figure~\ref{fig:cc_subjects}).
Apart from Subjects 13 and 15 (and to a lesser degree Subject 1) who appear to have different rating patterns from the majority of the subjects, the figure shows weak to moderate positive correlations between all subjects.
This suggests that in spite of different preferred rating ranges, the subjects made their judgments according to common criteria.

\newpage

\section{Prediction of optimal attenuation factors}\label{sec:learn-optim-atten}
In this section we describe the design and experimental validation of two modeling approaches to predict optimal attenuation factors from audio.
The models are ultimately intended to be used in a real-time plugin for audio workstations.
Although the details the real-time aspects of the implementation are beyond the scope of this paper, it does guide some important design decisions for the modeling.
Most importantly it implies a causal design in which the track cannot be analyzed as a whole in order to estimate the optimal attenuation factor.
On the other hand, the audio latency upper-bound for real-time operation (maximum observed audio latency in real-time commercial plug-ins is 4096 samples) is too low for accurate prediction of the attenuation factor.
This implies that the attenuation factor that will be applied at time~$t$ will be estimated from a windowed part of the signal (immediately) before $t$.
Whether the predicted attenuation factor is still approximately valid for the signal at time $t$ depends on the length of the window in relation to how quickly the resonance characteristics of the signal can change.
Our point of view is to target phenomena whose time-scale is around three seconds or longer---the same time scale as the \emph{short-term loudness} in EBU R128~\cite{ebur128}.
From this perspective we consider a window size of 0.5s a good trade-off, offering sufficient data for an informed prediction and at the same time being short enough to adapt to changes in resonance characteristics at the 3s time-scale.

We describe and test two alternative neural network approaches to the problem of predicting optimal attenuation factors.
The first is a more traditional approach in which a feature set is computed from an audio window, from which the attenuation factor is predicted.
The second approach skips the intermediate feature representation.
Instead, it takes the stereo PCM signal directly as input to a neural network that predicts the attenuation factor.

\subsection{Predicting based on audio features (FFN)}\label{sec:pred-based-audio}
Performing regression or classification tasks on audio using feature descriptions of the audio has been the predominant approach for the past decades,
and is based on the intuition that the prediction is determined by characteristics of the signal that can be defined explicitly and computed from the audio.
These descriptors often capture spectral characteristics of the signal, but may also approximate perceptual characteristics, such as loudness.
Many audio descriptors that have been proposed in the literature over time are implemented in a software library called Essentia~\cite{BogdanovEtAl_2013_EsseAnAudiAnal}.
The descriptors used in this study are listed in Table~\ref{tab:features}.
All descriptors are available in the Essentia library, except \emph{harmonics-to-noise ratio}~\cite{Boersma93accurateshort-term} and \emph{stereo width} (two descriptors, computed as the correlation between channels and absolute difference in RMS between channels, respectively), for which we used our own implementation.

The features are computed on shorter timescales (typically 1024 samples) than the 0.5s audio window for which our prediction will be made.
Thus the feature computation stage returns a vector of values for each feature.
We summarize each of these vectors by 7 statistics: the \emph{mean}, \emph{median}, \emph{standard deviation}, \emph{skew}, \emph{kurtosis}, the \emph{$10^{th}$ percentile}, and the \emph{$90^{th}$ percentile}.
This yields a total of 679 values per data instance, based on which a prediction must be made.
A common preprocessing step for high-dimensional input spaces like this is dimension reduction by means of \emph{principal component analysis} (PCA).
Tests showed that PCA did not lead to any substantial improvements however, and was thus not included in the final experiment.

\begin{table}
\caption{List of audio descriptors used in the FFN.}
\label{tab:features}
\small

\rule{\linewidth}{1pt}
\centering

\vskip1ex
\begin{vwcol}[widths={0.4,0.6},
sep=.8cm, justify=center, rule=0pt,indent=0em]
MFCC (13 values) \\
GFCC (13 values) \\
inharmonicity \\
pitch \\
pitch salience \\
spectral complexity \\
spectral crest \\
spectral decrease \\
spectral energy \\
spectral flux \\
spectral rms \\
spectral rolloff \\
spectral strong peak \\
zero-crossing rate \\
spectral flatness dB \\
high frequency coefficient \\
barkbands (30 values) \\
pitch instantaneous confidence \\
silence rate (at 20/30/60dB) \\
odd-to-even harmonic energy ratio \\
spectral energy band (4 values) \\
tristimulus (3 values) \\
spectral contrast (6 values) \\
spectral valley (6 values) \\
stereo width (2 values) \\
harmonics-to-noise ratio \\
\end{vwcol}

\rule{\linewidth}{1pt}
\end{table}

The network consists in a stack of linear layers (also called \emph{dense}, or \emph{fully connected}), each of which is followed by a \emph{batch normalization} (BN) layer and a layer of \emph{rectified linear} units (ReLU).
The BN layer transforms the distribution of the output activations of the preceding linear layer to zero mean and unit variance by keeping track of mean and variance during the training of the model.
The ReLU layer performs a non-linear transformation by setting negative output activations of the preceding layer to zero.
The number of linear layers and their sizes are not fixed in advance but determined using a hyper-parameter optimization scheme (Section~\ref{sec:experiments}).
A final linear layer is added after the after the last ReLU layer.
This layer has a single output---the predicted resonance attenuation factor.

\subsection{Predicting directly from audio: Dilated Residual Networks (DRN)}\label{sec:pred-directly-from}
In this section we describe a convolutional neural network that takes slices of a stereo PCM signal of the audio as input and provides an estimate of the optimal \attfact.
Note that even for a window of moderate size and sample rate this quickly leads to tens of thousands of samples to  be taken as model inputs.
As opposed to a feature vector however, the inputs are ordered along a meaningful dimension (time), in which patterns can be identified.
A common way do deal with data exhibiting such topological properties (sound, images, video) in neural networks is to use \emph{convolution}.
This approach, which was pioneered in~\cite{lecun98gradient-basedlearning}, exploits the fact that such data display local patterns that may occur at different locations in the data.
The strength of convolutional networks is that they learn to recognize patterns independently of their absolute location, and at the same time the convolution operation is much more space efficient than the ``fully-connected'' matrix dot product that takes place in regular feed-forward neural networks, allowing for larger models.
By stacking convolutional layers on top of each other it is possible to detect patterns of increasing size, and by interleaving the convolution operation with so-called \emph{pooling} or \emph{sub-sampling} operations, the patterns become somewhat invariant to local deformations.

However promising, the potential of traditional convolutional networks has been limited by a number of factors.
Two of these limitations have been addressed by recent extensions of the traditional convolutional network approach, namely \emph{dilated convolution}, and \emph{residual networks}.
We integrate both extensions in our convolutional network for predicting resonance attenuation factors, and discuss each of them briefly before we describe the global architecture of the model.

\subsubsection{Dilated convolution}\label{sec:dilated-convolution}
An approach often used with traditional convolution in order to create high-level feature representations of data is pooling using \emph{max} or \emph{average} aggregation functions.
For instance, max-pooling sub-samples the input by selecting maximal elements in a sliding window, typically using a sliding step (\emph{stride}) equal to the size of the pooling window.
Stacking convolution/pooling operations leads to features with increasing \emph{receptive fields}, meaning that the features can describe patterns of increasing size.
However, it comes at the cost of resolution loss: The relative position of features becomes less precise as their size increases.

On the contrary, dilated convolution achieves high-level features without loss of resolution.
Rather than increasing stride, it increases the receptive field of the features by ``dilating'' the convolution kernels.
A normal convolution of the kernel $k$ with the signal $s$ involves multiplying kernel elements with contiguous signal samples ($\tau$ is a discrete variable that increases in steps of 1):

\begin{equation}
\label{eq:conv}
(k * s)(t) = \sum_{\tau = -\infty}^{\infty} k(\tau)\ s(t - \tau)
\end{equation}

In convolution with dilation factor $d \in \mathbb{Z}^{+}$ on the other hand the kernel elements are multiplied with signal samples that are equally spaced at $d$ samples:

\begin{equation}
\label{eq:dilconv}
(k *_d s)(t) = \sum_{\tau = -\infty}^{\infty} k(\tau)\ s(t - d \tau)
\end{equation}

By stacking convolutional layers with increasing dilation factors the higher level filter kernels aggregate information over input ranges of exponentially increasing size, even if the size the kernels (in terms of parameters) does not increase, and the resolution remains intact.
This approach has proven successful in image processing tasks such as semantic segmentation~\cite{DBLP:journals/corr/YuK15}.

\newpage

\subsubsection{Residual blocks}\label{sec:residual-blocks}
Another issue with convolutional networks is that as they grow deeper in order to capture higher level patterns, it becomes harder to optimize the lower level convolutional layers.
This is directly related to the fact that for low level feature activations to influence the output of the model, they must pass through multiple layers of convolutions.
Sometimes however, it is desirable for low level features to be able to directly influence the output of the model, not just to figure as a building block for higher level features.

\begin{figure}
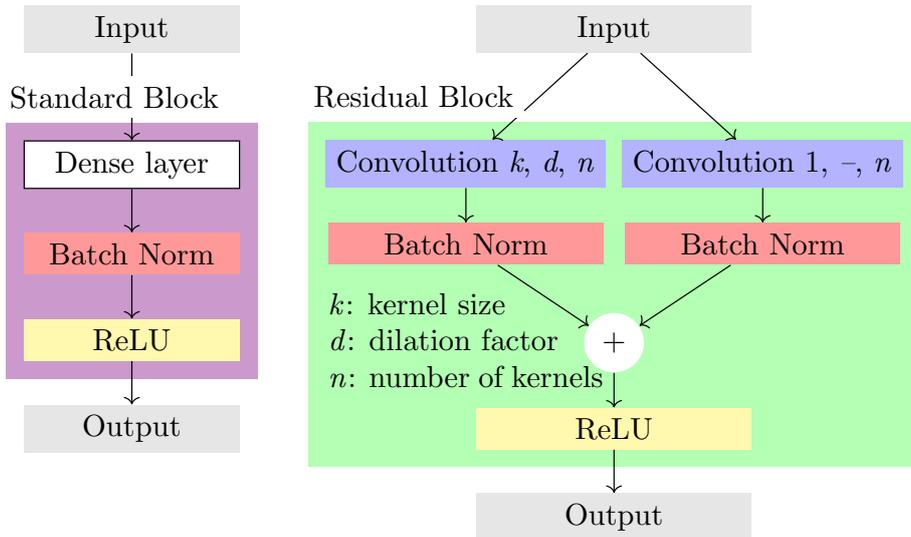

\centering
\begin{tikzpicture}[
  node distance = .5cm
  ]
  \input{figs/nn-preamble}
  \renewcommand{\mins}{7em}
  
  \node[data] (input) {Input};
  \node[default, below = 1cm of input](dense)  {Dense layer};
  \node[bn,   below = of dense] (bn)     {Batch Norm};
  \node[relu, below = of bn] (relu)   {ReLU};
  \node[data, below = of relu] (output) {Output};

  \foreach \from/\to in {
    dense/bn,
    bn/relu,
    relu/output}
  \draw [->] (\from) -- (\to);
  \draw [->] (input) --node[fill=white,xshift=-.2cm] {Standard Block} (dense);

  \begin{scope}[on background layer]
  \node[inner sep=2mm, fill=violet!40,
  fit=(dense) (dense) (relu) (relu)] {};
  \end{scope}
\end{tikzpicture}
\hfill
\begin{tikzpicture}[
  node distance = .4cm
  ]
  \input{figs/nn-preamble}
  \renewcommand{\mins}{9em}
  
  \node[data] (input) {Input};
  \node[conv, below left = 1cm and -1.5cm of input] (conv1) {Convolution \emph{k}, \emph{d}, \emph{n}};
  \node[conv, below right = 1cm and -1.5cm of input] (conv2) {Convolution $1$, \emph{--}, \emph{n}};
  \node[bn,   below = of conv1] (bn1)   {Batch Norm};
  \node[bn,   below = of conv2] (bn2)   {Batch Norm};
  \node[plus, below = 3cm of input]   (plus1) {+};
  \node[relu, below = of plus1]   (relu2) {ReLU};
  \node[data, below = .5cm of relu2] (output) {Output};
  \node[below = .2cm and -3cm of bn1] (legend) {
    \parbox{9em}{
    \emph{k}: kernel size\\
    \emph{d}: dilation factor\\
    \emph{n}: number of kernels\\
    }
  };

  \foreach \from/\to in {
    conv1/bn1,
    bn1/plus1,
    input/conv2,
    conv2/bn2,
    bn2/plus1,
    plus1/relu2,
    relu2/output}
  \draw [->] (\from) -- (\to);
  \draw [->] (input) --node[fill=white,xshift=-1.45cm] {Residual Block} (conv1);

  \begin{scope}[on background layer]
  \node[inner sep=2mm, fill=green!30,
  fit=(conv1) (relu2) (relu2) (conv2)] {};
  \end{scope}
\end{tikzpicture}
\caption{Building blocks for the FFN and DRN models.
Left: Standard block composed of a dense linear layer followed by batch normalization and a rectified-linear layer (See Section~\ref{sec:pred-based-audio}); Right: Residual block (See Section~\ref{sec:residual-blocks}).}
\label{fig:resblock}
\end{figure}

This observation has led to the proposal of the \emph{residual block} as a sub-structure used in deep networks~\cite{he16:_deep_resid_learn_image_recog,DBLP:journals/corr/YuKF17}, an adaptation of which is depicted in Figure~\ref{fig:resblock}.
In this structure the information flows from input to output through two pathways in parallel.
The left pathway involves a typical convolution layer with configurable parameters: the kernel size $k$, number of kernels $n$, and the dilation factor $d$.
The convolution in the right pathway uses kernels of size one (the dilation factor is irrelevant in that case), and thus does not compute any features from the input.
Instead, it outputs $n$ linear combinations of the input in order to make the input shape compatible for elementwise addition to the $n$ \emph{feature maps} of the left pathway.
Both pathways further include a batch normalization operation.
After the elementwise sum of both pathways a rectified linear unit allows for a non-linear response.

The term ``residual'' refers to the fact that the left pathway only needs to account for the part of the output that cannot be accounted for by linear combinations of the input---the right pathway.
In this way, increasing the number of stacked convolution operations does not hamper the ability of the network to account for its output in terms of lower level features.

\begin{figure}[!h]
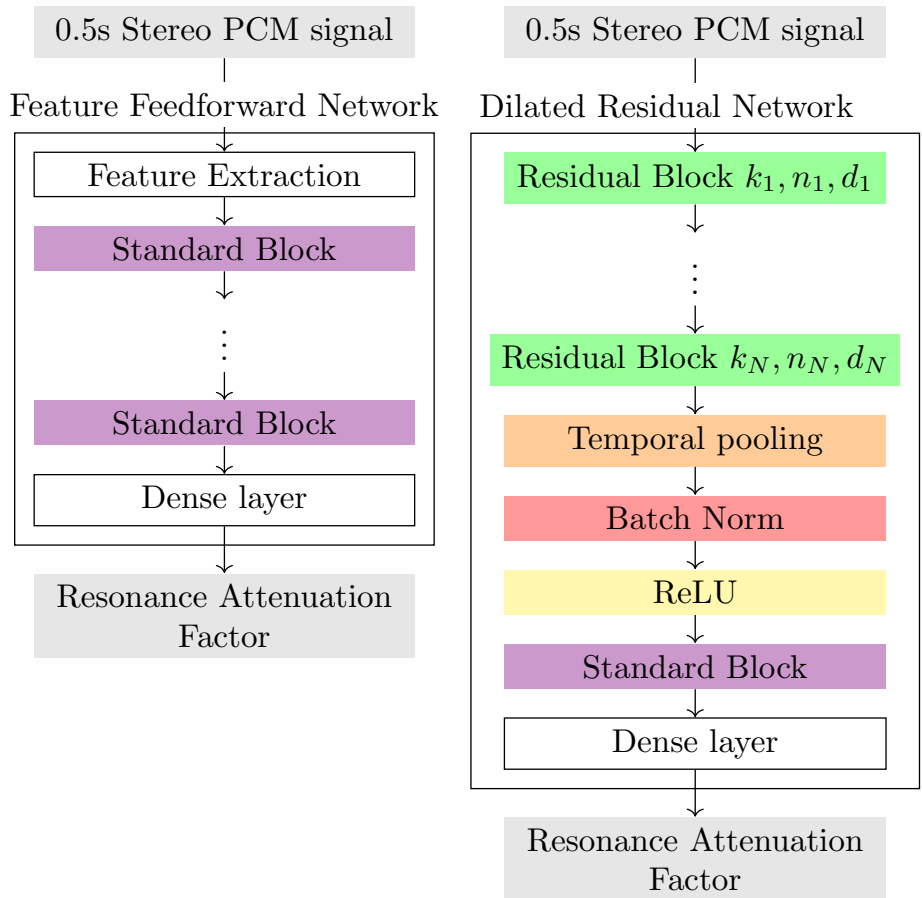

\centering
\begin{tikzpicture}[
  node distance = .3cm
  ]
  \input{figs/nn-preamble}
  \renewcommand{\mins}{11.5em}

  \node[data] (input) {0.5s Stereo PCM signal};
  \node[default, below=1cm of input] (feat) {Feature Extraction};
  \node[sb, below= of feat] (sb1) {Standard Block};
  \node[below= of sb1] (ell) {$\vdots$};
  \node[sb, below= of ell] (sb2) {Standard Block};
  \node[default, below= of sb2] (dl2) {Dense layer};
  \node[data, align=center, below= .5cm of dl2] (output) {Resonance Attenuation\\ Factor};

  \foreach \from/\to in {
    feat/sb1,
    sb1/ell,
    ell/sb2,
    sb2/dl2,
    dl2/output}
  \draw [->] (\from) -- (\to);
  \draw [->] (input) --node[fill=white,xshift=0cm] {Feature Feedforward Network} (feat);

  \begin{scope}[on background layer]
  \node[draw,inner sep=2mm,
  fit=(feat) (dl2) (sb1) (dl2)] {};
  \end{scope}
\end{tikzpicture}
\hfill
\begin{tikzpicture}[
  node distance = .3cm
  ]
  \input{figs/nn-preamble}
  \renewcommand{\mins}{11.5em}

  \node[data] (input) {0.5s Stereo PCM signal};
  \node[rb, below=1cm of input] (rb1) {Residual Block $k_1, n_1, d_1$};
  \node[below= of rb1] (ell) {$\vdots$};
  \node[rb, below= of ell] (rb4) {Residual Block $k_N, n_N, d_N$};
  \node[pool, below= of rb4] (pool) {Temporal pooling};
  \node[bn, below= of pool] (bn1) {Batch Norm};
  \node[relu, below= of bn1] (relu1) {ReLU};

  \node[sb, below= of relu1] (sb) {Standard Block};
  
  \node[default, below= of sb] (dl2) {Dense layer};
  \node[data, align=center, below= .5cm of dl2] (output) {Resonance Attenuation\\Factor};

  \foreach \from/\to in {
    rb1/ell,
    ell/rb4,
    rb4/pool,
    pool/bn1,
    bn1/relu1,
    relu1/sb,
    sb/dl2,
    dl2/output}
  \draw [->] (\from) -- (\to);
  \draw [->] (input) --node[fill=white,xshift=-.3cm] {Dilated Residual Network} (rb1);

  \begin{scope}[on background layer]
  \node[draw,inner sep=2mm,
  fit=(rb1) (dl2) (rb4) (rb4)] {};
  \end{scope}
\end{tikzpicture}
\caption{FFN and DRN architectures.}
\label{fig:model}
\end{figure}

\vspace{.6cm}

\subsubsection{DRN architecture}\label{sec:drn-architecture}
Figure~\ref{fig:resblock} shows the complete model consisting of multiple residual blocks.
Note that the residual blocks maintain the original temporal dimension of the data, which amounts to a size of 11025 for 0.5s of audio sampled at 22050Hz.
The temporal pooling operation reduces this number by down-sampling the output using window-wise averaging, and is followed by two dense layers with intermediate batch-normalization and non-linearity in order to produce an estimated resonance attenuation factor.

\newpage

\subsection{Experiments}\label{sec:experiments}
In this section we describe the training and evaluation procedure of both model architectures described above.
We use the human ratings of the 150 tracks gathered in the listening experiment to train and evaluate both architectures.

\subsubsection{Procedure}\label{sec:procedure}

\paragraph*{Evaluation Criterion}
Predicting optimal resonance attenuation factors for given tracks is a regression problem and as such an obvious choice is to use the mean squared error of the predictions with respect to the optimal value (the \emph{target}) as an objective to be minimized.
However, given the variance in the ratings across subjects in the ground truth, it may be hard to determine a unique optimal value per track.
Using the mean or median of the ratings per track as a target has the drawback that the mean squared error objective does not differentiate between tracks with different degrees of rater consensus.
Ideally, we wish to impose a lower penalty on errors from the mean rating when the rater consensus is low.
We do so by generalizing the mean squared error objective as follows.
Rather than defining the objective function to be minimal only for a single value, we define it to be minimal whenever the prediction lies within a specified interval that varies from one data instance to another.
For a given data instance consisting of an audio track $A \in \mathbf{A}$ and a set of ratings $\mathbf{\af}$ we define the zero penalty interval as $[\ P_{\mathit{l}}(\mathbf{\af}),\ P_{\mathit{h}}(\mathbf{\af})\ ]$, where $P_l(\mathbf{\af})$ and $P_h(\mathbf{\af})$ denote the $l$-th and $h$-th percentiles of the ratings $\mathbf{\af}$, respectively, with $l \leq h$.
We refer to this objective as the \emph{mean squared bounds errror} with bounds $\mathit{l}$, $\mathit{h}$, or \msbe{l}{h}.
We use $l=35$ and $h=65$ throughout the experiments.

Formally, given a dataset $D$ consisting of pairs $(A, \mathbf{\af})$, the \msbe{l}{h} of a model $f: \mathbf{A} \rightarrow \mathbb{R}$ is defined as:

\begin{equation}
\label{eq:boundsloss1}
\msbe{l}{h}(f, D) = \frac{1}{\left|D\right|}\sum_{(A,\ \mathbf{\af})\ \in\ D} L_{A,\mathbf{\af}}^{(l,\, h)}(f),
\end{equation}

where

\begin{eqnarray}
\label{eq:boundsloss2}\nonumber
L_{A,\mathbf{\af}}^{(l,\, h)}(f)  & = & \left( \left[\ f(A) - P_{\mathit{h}}(\mathbf{\af})\ \right]^+ \right.
+ \\
&   & \left.
\vphantom{\left[ f(A) - P_{\mathit{h}}(\mathbf{\af})\ \right]^+}\ \ \left[\ f(A) - P_{\mathit{l}}(\mathbf{\af})\ \right]^{-}  \right)^2.
\end{eqnarray}

The brackets $\left[\ \cdot\   \right]^{+}$ and $\left[\ \cdot\ \right]^{-}$ denote the positive and negative parts respectively.

\newpage

\paragraph*{Hyper-parameter optimization}
We use Bayesian optimization to find the optimal hyper-parameters for each of the models, most importantly the depth of the networks and the hidden layer sizes.
This is a heuristic to speed up the search for appropriate hyper-parameter values compared to an exhaustive grid search.
The particular form of optimization we use is based on a gaussian process approximation of the loss as a function of the hyper-parameters.
This approximation gives rise to the \emph{upper confidence bound}~\cite{Lai:1985:AEA:2609660.2609757}, which estimates the expected loss for hyper-parameter settings that have not yet been tested, and is used as a guide to search the space of hyper-parameters~\cite{snoek2012practical}.

Apart from the depth of the models and the hidden layer sizes, the optimization involved hyper-parameters to control the training procedure: the \emph{learning rate}, and the thresholds for \emph{early stopping}, and \emph{learning rate reduction}.

\paragraph*{Cross-validation}
To perform the hyper-parameter optimization we use two partitions of the dataset into a test set (10 tracks), a validation set (10 tracks), and a train set (130 tracks).
For each of the test tracks we compute the \msbe{35}{65} loss on 100 randomly selected 0.5s frames.
The criterion used to optimize the hyper-parameters is the average frame-wise loss across both test sets.

With the best hyper-parameters found for the FFN and DRN architectures, respectively, we perform a further five fold cross-validation.
To this end, we use the same dataset, but exclude the 20 test tracks used for hyper-parameter optimization.
We repeat the five fold cross validation five times using different random seeds to reduce the effect of partitioning of the data into folds and model parameter initializations on the result.

\paragraph*{Baseline} We define a baseline approach as a reference for evaluating the FFN and DRN architectures.
This approach consists in computing the mean resonance attenuation factor observed over all tracks in the training set, and using this value as a prediction for the test set, irrespective of the input audio.

\subsubsection{Results and discussion}
\begin{figure}
\begin{center}
\includegraphics[width=.8\linewidth]{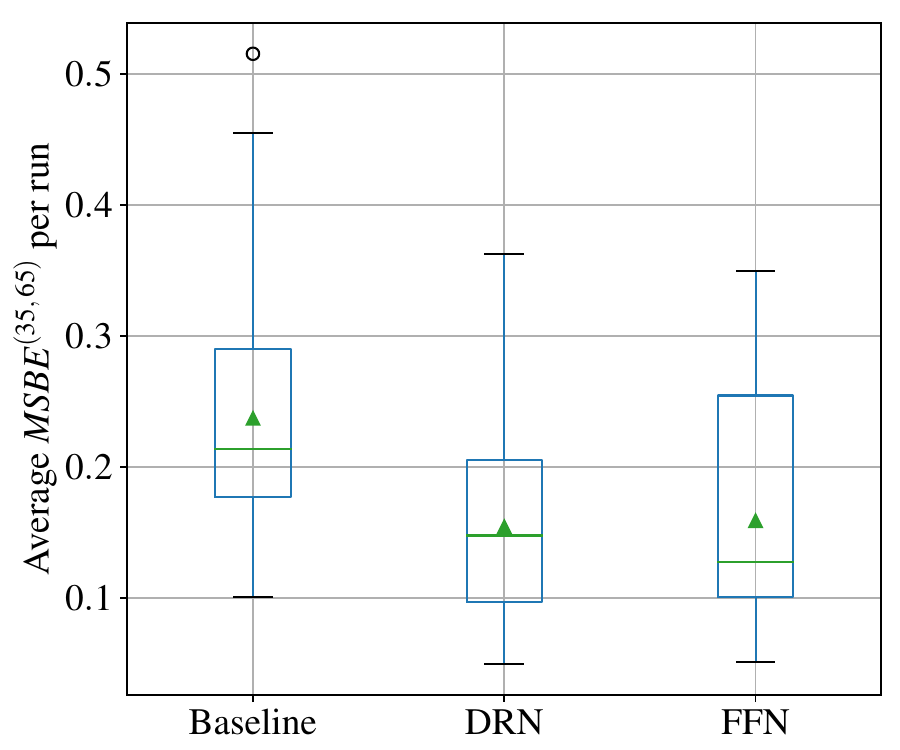}
\end{center}
\caption{Average \msbe{35}{65} per run for baseline, FFN, and DRN models; The horizontal lines in the boxes indicate the median, triangles the mean values per model.}
\label{fig:box_model}
\end{figure}

The optimal configuration for the FFN and the DRN architectures, as found by hyper-parameter optimization, are shown in Table~\ref{tab:hpars}.
Figure~\ref{fig:box_model} shows the results of these architectures on the repeated five fold cross validations.
A one-way repeated measures ANOVA reveals a significant effect of model on \msbe{35}{65}
($F_{2,\,72}=6.55$, $p=0.002$).
A post-hoc Tukey HSD test at $\alpha = 0.05$ indicates that DRN and FFN differ significantly from the baseline.
The effect size of DRN over baseline corresponds to Cohen's $d=0.88$, whereas the FFN over baseline effect size is $d=0.82$.
The difference between DRN and FFN is not significant.

\begin{table}
\caption{Optimal configuration for the FFN and the DRN architectures as found by hyper-parameter optimization.}
\label{tab:hpars}
\small
\centering
\begin{tabularx}{\linewidth}{l>{\centering\arraybackslash}X>{\centering\arraybackslash}X}
\toprule
& FFN & DRN \\
\midrule
Depth & 3 Std. Blocks & 10 Res. Blocks \\
Block Size & & \\
\hspace{1em}(Low/Mid/High) & 500 / 250 / 250 & 100 / 100 / 300 \\
Temporal Pooling & -- & 300 \\
Final Std. Block Size & -- & 10 \\
\bottomrule
\end{tabularx}
\end{table}

\begin{table}
\caption{Means, standard deviations, and the 95\% confidence interval (CI) for the mean \msbe{35}{65} per model.}
\label{tab:groupstats}
\small
\centering
\sisetup{
round-mode      = places,
round-precision = 3,
table-format=1.3
}
\begin{tabular}{lSSSS}
\toprule
&                     &                      & \multicolumn{2}{c}{95\% CI} \\
\cmidrule(lr){4-5}
Model    & {Mean}              & {Std.
dev.}          & {Low}               & {High}       \\
\midrule
Baseline & 0.23713116796410347 & 0.10264326127224099  & 0.19388843140545128 & 0.28037390452275562 \\
FFN      & 0.15907279158723805 & 0.082345795158922999 & 0.12438120557726541 & 0.19376437759721071 \\
DRN      & 0.15443630659215482 & 0.079993701630128367 & 0.12073563766301429 & 0.18813697552129535 \\
\bottomrule
\end{tabular}
\end{table}

Table~\ref{tab:hpars} shows that the FFN architecture works best when it is comparatively shallow (three standard blocks, the minimal depth tested), whereas the DRN architecture performs better when it is deep (10 residual blocks, the maximal depth tested).
This trend is consistent in a review of the 10 best FFN and DRN architectures as found in the hyper-parameter optimization, omitted here for the sake of brevity.
The layer sizes however do not show a similarly consistent trend, and vary considerably throughout the 10 best FFN and DRN architectures.

The fact that both modeling approaches show similar accuracies on the attenuation factor prediction
is in line with a general trend that arises from the deep learning literature, showing that current techniques in deep learning 
are powerful enough to work with raw digitizations of information---such as sampled waveforms---and still address tasks that require considerable abstraction from that low level representation, thus reducing the need for hand designed features.

At the same time, it must be noted that the roughly equivalent performance of the FFN and DRN measured here is at odds with a multitude of cases where end-to-end deep networks clearly outperform prior state-of-the-art methods that rely on a hand-designed feature extraction stage, especially in image processing~\cite{zheng2018sift}.
For audio tasks such as automatic tagging however, end-to-end networks do not seem to have a strong advantage over spectrogram-based approaches~\cite{dieleman2014end}, and require large training data sets in order to outperform spectrogram-based approaches in audio tagging tasks~\cite{pons17:_end}.
Similarly, in the study presented here the small size of the data set---especially in combination with inter-subject variance and the non-uniform distribution of the ratings---is a plausible explanation of why the DRN does not outperform the FFN.

\section{Conclusion}\label{sec:conclusion}
In this paper we addressed the problem of automatic resonance equalization.
We have proposed a method to attenuate automatically identified resonances by a user-controlled factor.
Using this method we carried out a listening experiment in which sound engineers identify optimal attenuation factors for a set of audio tracks.
The results, which show general consensus in ratings among subjects, were used to train and evaluate two types of predictive models to estimate optimal resonance attenuation factors based on the content of the audio tracks.
The results show that an intermediate stage of feature extraction is not strictly necessary for this task: a dilated residual network performs equally well when applied directly to the audio signal.

The proposed system is a fully auto-adaptive resonance equalization system in which the attenuation factor is chosen automatically by a deep neural network.
To our knowledge, this system is the first documented self-adaptive equalizer based on neural networks.
Future work includes a real-time implementation of the presented model as a real-time plugin that can be used in audio work stations.

\newpage

\bibliographystyle{plain}
\bibliography{references}

\end{document}